\documentclass[prl,twocolumn,superscriptaddress,floatfix]{revtex4}
\usepackage{amsmath,amssymb,graphicx}
\begin{document}
\title{Absence of Critical Thickness in an Ultrathin Improper Ferroelectric Film}
\author{Na Sai}
\affiliation{Department of Physics, The University of Texas at Austin, Texas, 78712}
\affiliation{Center for Computational Materials, Institute for Computational Engineering and
Sciences, The University of Texas, Austin, TX 78712}
\author{Craig J. Fennie}
\affiliation{School of Applied and Engineering Physics, Cornell University, Ithaca NY, 14853}
\author{Alexander A. Demkov}
\affiliation{Department of Physics, The University of Texas at Austin, Texas, 78712}
\date{August 19, 2008}

\begin{abstract}
We study the ferroelectric stability and surface structural properties of an oxygen-terminated hexagonal YMnO$_3$ ultra-thin film using density functional theory. Under an open circuit boundary condition, the ferroelectric state with the spontaneous polarization normal to the (0001) surface, is found to be metastable in a single domain state despite the presence of a depolarizing field. We establish a connection between the result and the role of improper ferroelectric transition. Our results imply that improper ferroelectric ultrathin films can have rather unique properties that are distinctive from those of very thin films of ordinary ferroelectrics. 
\end{abstract}
\maketitle

The integration of ferroelectric oxide materials into existing microelectronic device architectures 
is currently of much interest as they hold promise for a wide range of potentially new 
applications~\cite{JimScott.science,Dawber05}. This has generated an enormous effort to 
understand the properties of ferroelectric perovskite ultrathin films, such as BaTiO$_3$,
where it has been shown that electrical boundary conditions~\cite{Ghosez,Meyer,Junquera,Kornev}, 
surface and interface properties~\cite{Sai05,Fong,Gerra,Sepliarsky}, as well as epitaxial 
strain~\cite{Tinte,Dieguez} all play an important role.  
 Although there appears to be no fundamental size limit below which ferroelectricity disappears, 
 ultrathin ferroelectric films with the polarization normal to the surface remain a challenge.
This is because the depolarization field arising from the accumulated charges at 
the surfaces, if not screened, can strongly suppress the instability towards a 
single-domain ferroelectric state. This is indeed what happens in a material displaying a proper ferroelectric transition (note, henceforth we refer to such materials as proper ferroelectrics), namely
one where the primary order parameter is the electrical polarization, e.g., BaTiO$_3$. The 
depolarizing field contributes to the free energy a positive term quadric in the polarization 
thereby re-normalizing the soft-mode energy and stabilizing the paraelectric phase.
Even in a system with metallic electrodes, which can provide the necessary  screening in 
most cases, when the thickness of the ferroelectric film becomes comparable to the effective 
screening length of the metal, the screening is incomplete resulting in a reduced 
polarization~\cite{Junquera,Sai05, Umeno}, an increased coercive field~\cite{Dawber03}, and
in some case suppress the tendency towards (single-domain) ferroelectricity completely.

It would clearly be advantageous, if not at least fundamentally interesting, to consider the
surface properties of ultrathin films of materials in which ferroelectricity did not originate
 from a polar instability, but rather from an improper ferroelectric transition -- i.e., one where 
the spontaneous polarization does not drive the transition but instead is a slave to some other primary
order parameter. As pointed out decades ago by Levanyuk and Sannikov in a series of papers~\cite{Levanyuk1, Levanyuk}, an instability towards a single-domain ferroelectric state in a improper ferroelectric 
is still possible even if the depolarization field remains unscreened, e.g., under open circuit boundary conditions.

\begin{figure}
\includegraphics[width = 6 cm]{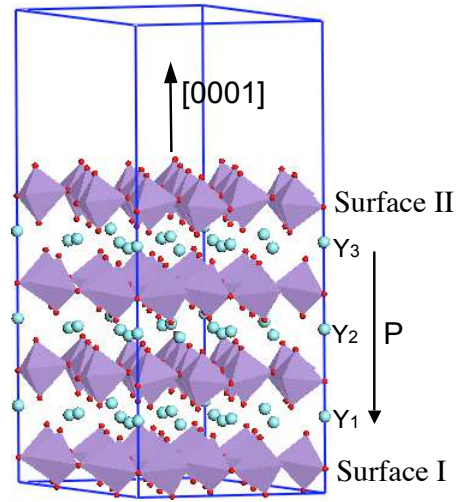}
\caption{Side view of YMnO$_3$ (0001) thin film slab two unit-cells thick symmetrically terminated with apical oxygens. The three Y (large blue circles) planes are indicated. Small red circles and polyhedra represent oxygens and MnO$_5$ cages. The polarization of the unrelaxed FE structure pointing from surface II to surface I is indicated.}
\label{fig:struc}
\end{figure}
In this letter we study for the first time finite size effects in an improper ferroelectric from first 
principles. We use hexagonal YMnO$_3$~\cite{Ismailzade, Lukas, Katsufuji, vanAken04p164}, 
which was recently predicted~\cite{Fennie05p100103} and subsequently experimentally 
confirmed~\cite{Namjung, Nenert.thesis} to display an improper ferroelectric transition as an 
example. We investigate the intrinsic properties of isolated thin films of YMnO$_3$ with (0001) 
orientation and a clean oxygen terminated surface {\it in the absence of an external field or 
electrode} and determine the influence of the surface on the ferroelectric state.
Despite the growth of YMnO$_3$ thin films on a wide range of substrates~\cite{Fujimura}, the structural properties of the surface have not been extensively studied previously.
We show that contrary to the case of a thin film proper ferroelectric, e.g., BaTiO$_3$ or 
PbTiO$_3$~\cite{Meyer, Sai05, Junquera}, the polarization persists in a single domain 
YMnO$_3$ film to approximately two unit cells thick (the smallest size in which a bulk region can be defined), i.e., the relaxation of the YMnO$_3$ slab {\it does not result in a paralectric structure}.

We carry out the first-principles study of YMnO$_3$ thin films using density functional theory as 
implemented in the VASP code~\cite{VASP} and projected augmented wave pseudopotentials. 
The Hubbard correction to the local density approximation (LDA+U) is applied to the Mn $3d$ 
orbitals to account for the correlation effect. We use $U = 6.0 ~{\rm eV}$ and $J = 0.9 ~{\rm eV}$ 
which was previously shown~\cite{Fennie05p100103,Sai07} to yield a band gap and lattice parameters 
close to the experiments. We start with a slab composed of  two unit cells, stacked along the hexagonal axis, of the bulk ferroelectric YMnO$_3$ structure, terminating symmetrically on the apical 
oxygens with a vacuum layer of $\sim10~{\rm\AA}$ added as shown in Fig.~\ref{fig:struc}. 
This is the minimum thickness for a YMnO$_3$ film for which a ``bulk'' region can be defined. 
We fully relax the atomic positions of the slab while keeping the in-plane lattice constants 
fixed at the theoretical bulk ferroelectric equilibrium values (i.e., $a = 6.09~{\rm \AA}$  and 
$c = 11.36~{\rm \AA}$). The planewave cutoff energy is 500 eV and a Monkhorst-Pack $4\times4\times1$ 
$k$-point mesh is used for the Brillouin zone integration. We have applied collinear approximation 
to describe the antiferromagnetic spin ordering while neglecting spin-orbital coupling and 
magnetic anisotropy, which was previously found  to be small compared to the structural 
energetics~\cite{Fennie05p100103}. 

Previous theoretical work has shown that the chemical bonding in rare-earth based hexagonal 
manganites such as YMnO$_3$ plays a very minor role in the origin of ferroelectricity~\cite{vanAken04p164}, 
unlike that of conventional perovskite ferroelectrics where covalency effects between the cations and 
oxygens are crucial~\cite{RonCohen}. The difference in the nature of bonding from that of perovskites also affects the geometric relaxation and surface band structure as we will show next.
\begin{figure}[htbp]
\includegraphics[width=8cm]{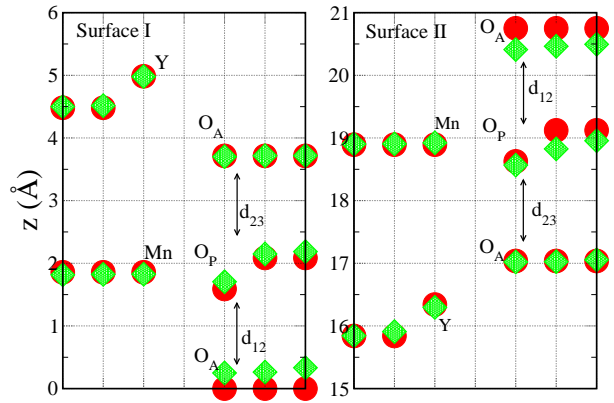}
\caption{Geometric relaxation of the bottom and top four surface atomic planes in the YMnO$_3$ slab along the surface normal direction. Circles and diamonds depict the unrelaxed and relaxed $z$-coordinates, respectively, for the surface MnO$_5$ and Y planes.}
\label{fig:sur_out}
\end{figure}

The geometric relaxation normal to the surface is illustrated in Fig.~\ref{fig:sur_out}, where I and II denote the same surfaces as in Fig.~\ref{fig:struc}. At both surfaces, the surface apical (O$_A$) oxygens sitting atop the MnO$_5$ cages move substantially inward relative to their unrelaxed positions, toward the bulk region, with a displacement ranging between $0.25~{\rm \AA}$ and $0.33~{\rm \AA}$. The displacements are larger than those reported for, e.g., BaTiO$_3$ surface~\cite{Padilla}, which measure of $\sim 0.1~{\rm \AA}$. This is unsurprising as there are no metal cations that bond with the apical oxygens in the common surface plane in YMnO$_3$. Instead, the surface oxygens bond strongly with the Mn atoms in the next atomic plane. The surface relaxation should thus be more significant than that found in perovskites. In the next layer down, the Mn atoms move towards the surface on average by $0.03~{\rm \AA}$ and $0.02~{\rm \AA}$, while the oxygens (O$_P$) move towards the bulk region by $0.08~{\rm \AA}$ and $0.17~{\rm \AA}$ for surfaces I and II, respectively. The apical oxygens (O$_A$) and Y layer below the surface plane have negligible relaxation ($\leq 0.01~{\rm \AA}$). Overall, the interplane distance (calculated from position averaged for each layer) at surface I between the first two surface planes $d_{12}$ reduces by $0.23~{\rm \AA}$ and between the second and third planes $d_{23}$ has reduced by $0.02{\rm \AA}$. Similarly, a reduction of $d_{12}$ by $0.21~{\rm \AA}$ and $d_{23}$ by $0.08~{\rm \AA}$ is found at surface II.

\begin{figure}[htbp]
\includegraphics[width = 8cm]{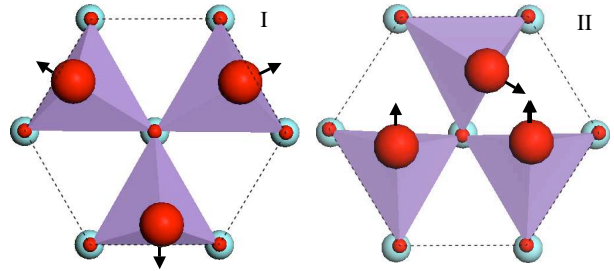}
\caption{Top view of the in-plane positions of the surface apical oxygens (large red circles) relative to the MnO$_5$ polyhedra for surfaces I and II. The blue medium circles depict Y atoms near the surface. The arrows depict the in-plane relaxation direction relative to the initial positions.}
\label{fig:sur_in}
\end{figure}
Surface I and II differ significantly with respect to their in-plane relaxation as shown in Fig.~\ref{fig:sur_in}. Three apical oxygens move outward along the in-plane unit-cell axis by about $0.38$, $0.4$, and $0.47~{\rm \AA}$ at surface I, displaying a tendency to compensate the shortened Mn-O$_A$ bond along the $c$ axis. At surface II, two of the three apical oxygens (in the unit cell) display a tendency to dimerize, yielding a O-O bond of $2.23~{\rm \AA}$, $3.14~{\rm \AA}$, and $3.59~{\rm \AA}$ (the length is $3.22~{\rm \AA}$, $3.67~{\rm \AA}$, and $3.67~{\rm \AA}$ in the bulk). The non-equivalence between the two surfaces evident from the in-plane relaxation arises from the presence of the ferroelectric polarization as we show below.

\begin{figure}
\includegraphics[width = 8cm]{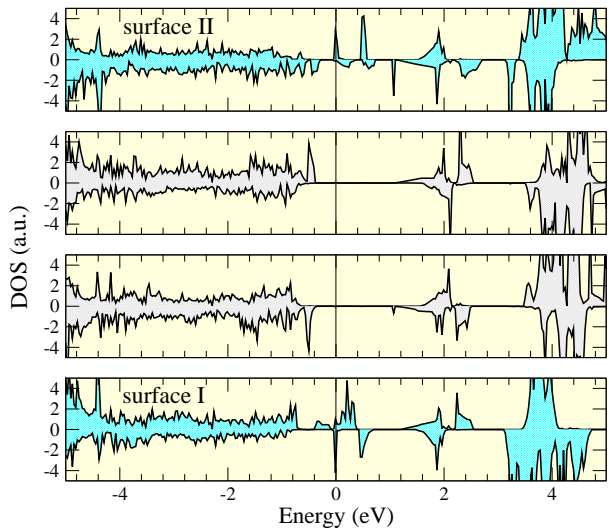}
\caption{The projected density of states for each MnO$_5$ plane along the YMnO$_3$ thin film $c$ axis as in Fig~\ref{fig:struc}. The top and bottom panels correspond to surface II and I. The polarization points from the top to the bottom surface. The Fermi level is marked at $E=0$.} 
\label{fig:dos_parchg}
\end{figure}
Fig.~\ref{fig:dos_parchg} shows the projected density of states (PDOS) for each MnO$_5$ plane along the $c$ axis of the film. Like the geometric relaxation, the electronic structure of the interior layers rapidly converges to that of bulk YMnO$_3$ where a band gap about 1.4 eV has been reported~\cite{Kang05p092405}. The top of the valence bands on the surfaces shows more changes than the conduction bands with respect to the bulk, as the valence band is partially composed of O 2p of apical oxygens lying in the surface plane. There are surface states that lie deep near the center of the gap which pin the Fermi level in the gap. These states are primarily of the O $2p$ character admixed with the Mn $3d$ and are localized at the surface plane of the oxygen terminated surface. This is somewhat different from a perovskite such as BaTiO$_3$, where surface states were found to intrude only into the lower part of the gap as a result of the strong hybridization between Ti and O~\cite{Padilla}. Quantitatively, the band structure of the YMnO$_3$ film relaxed in the FE structure looks similar to that obtained for a paraelectric structure (not shown), however, one can notice a slight asymmetry in the relaxed FE structure between the two surface states, namely, a slight excess of electrons at surface I and excess of holes at surface II. As we will show the asymmetry can be traced to the polarization across the film normal to its surface. 

By determining the potential drop across the fully relaxed YMnO$_3$ film in our calculation, we 
evaluate the depolarizing field $E_d$ of $0.025~{\rm eV/\AA}$ across the 
film~\cite{dip}. This electric field in the YMnO$_3$ slab is approximately an order of magnitude
smaller than that of an isolated PbTiO$_3$ slab~\cite{Sai05} (the latter can be calculated by 
artificially imposing a constraint on the ions so that the structure does not relax back to a 
paraelectric state) and is of course consistent with the fact that the ferroelectric Berry phase 
polarization~\cite{KingSmith} in bulk YMnO$_3$ of $6.5~\mu{\rm C/cm}^2$ is approximately 
one order of magnitude smaller than the bulk polarization in PbTiO$_3$. 

The magnitude of the field in  a YMnO$_3$ film appears to indicate that not only does a spontaneous 
polarization exist in an isolated slab of YMnO$_3$ but that the value is comparable to that in the bulk.  
Next we calculate, albeit approximately, the polarization in the YMnO$_3$ slab via ionic 
displacements from a paraelectric reference structure to the relaxed ferroelectric structure, ${\bf u}$, 
and Born effective charges, ${\bf Z}$, where $P\approx~{\bf Z}\cdot{\bf u}$~\cite{Pmethod}. The resulting 
polarization equals to $\sim6.1~\mu$C/cm$^2$ which is only slightly smaller than the bulk polarization. 

So why does an uncompensated single domain slab of BaTiO$_3$ (or PbTiO$_3$ for that 
matter) remains paraelectric yet a slab of YMnO$_3$ under similar conditions becomes 
ferroelectric? The simplest way to understand this difference is to investigate a phenomenological 
free energy.
In materials that display a proper ferroelectric transition such as BaTiO$_3$, the free energy
can be expanded in terms of the polarization: 
$$\mathcal{F}(P) = 1/2\alpha P^2 + 1/4 \beta P^4 + \mathcal{O}(P^6)$$
where the ferroelectric transition occurs when $\alpha$ =0 (for purposes of illustration we 
are only discussing second order phase transitions). In the case of an uncompensated film a
depolarizing field, $\vec{E}_d\propto -{\vec P}$, adds a positive term to the free energy, 
$\mathcal{F}_d\sim P^2$, renormalizing $\alpha $ to $\alpha'$, where due to the 
comparatively large depolarization energy, $\alpha'$ $>$ 0 at all temperatures, suppressing 
the ferroelectric transition into a single domain state. The polarization charges on the surface 
must be compensated by free charges provided by, e.g., external electrodes, to preserve the 
polarization. 

In contrast, YMnO$_3$ was shown to display an improper ferroelectric transition where the 
polarization arises due to a non-trivial coupling to a zone-boundary lattice instability, the $K$-mode. 
A simplified free energy is given by:
$$\mathcal{F}(P,K) = \alpha_{02}P^2  + \alpha_{20}K^2 + \beta_{40}K^4 + \beta_{31}K^3P + ...$$
where $K$ corresponds to the primary order parameter associated with the zone boundary $K_3$ 
phonon mode (accounting for the transition $\sim$1200K) and $P$ is the 
polarization~\cite{Fennie05p100103}. Here the quadratic coefficient of the polarization, $\alpha_{02}$,
does not soften to zero at the ferroelectric transition~\cite{coeff}, i.e., $\alpha_{02}$ $>$ 0 at any temperature, 
and  the spontaneous polarization arises because of the $\beta_{31}$ coupling, where for small 
$K$, $P$$\sim$ $K$$^{3}$. Physically this coupling acts like a field, where $K_3$ ``pushes'' the 
equilibrium value of the 
single minimum potential well of the polar mode to nonzero value. 
The energy associated with this field, as previously calculated in bulk YMnO$_3$ 
($\sim5\times 10^{-4}~{\rm eV/\AA}^3$, or $0.06~{\rm eV}$ per formula unit), turns out roughly 6 
times larger than the depolarizing energy. For an uncompensated film in this situation,
the effect of the depolarization field is 
to stiffen the already "hard" quadratic coefficient $\alpha_{02}$, which can effectively lower the equilibrium value of 
the polarization (see Ref.~\onlinecite{Fennie05p100103}) but in principle does not suppress the 
instability to a single domain ferroelectric state. 

We have studied the ferroelectricity and surface relaxation of oxygen terminated YMnO$_3$ thin films. 
Using ultrathin films of YMnO$_3$ as an example, we show that in improper ferroelectrics, the 
instability to a single domain ferroelectric state does not vanish even when the depolarization field 
remains unscreened, i.e., in the absence of electrodes or external field. Our results are applicable for
all materials that display an improper ferroelectric transition, whether the polarization is induced 
by structural changes, like the case we considered here, or by for example a spin ordering. This has potential exciting applications given that a wide class of perovskite 
nanostructures have been recently found to display improper ferroelectricity~\cite{Bousquet}. 

This work was supported by the Office of Naval Research grant N000 14-06-1-0362 and National Science Foundation under grant DMR-0548182. The computing was done at the Texas Advanced Computing Center (TACC) at the University of Texas at Austin.


\begin{thebibliography}{10}
\bibitem{dip}There is also a small field of $0.0056~{\rm eV/\AA}$ in the vacuum region pointing in the opposite direction arising from the periodic boundary condition applied to the supercell. Note that we did not apply an external field~\cite{Meyer} because we do not intend to simulate a situation of ``idealized electrodes'' or a short circuit condition. Instead we have applied a dipole correction~\cite{Bertgsson} which removes the field in the vacuum region by introducing a small potential discontinuity.
\bibitem{Pmethod} First we calculate the lattice parameter $c$ along the hexagonal axis from the separation between the average position of the Y$_1$ and Y$_3$ planes (see Fig.~\ref{fig:struc}), and thus determine the complete paraelectric atomic coordinates from the symmetry of the P6$_{\rm 3}$/mmc group. The polarization for the central unit cell of the fully relaxed thin film structure is then calculated from the displacements of the planary oxygens and the Y atoms in the midplane relative to their paraelectric positions, and from the corresponding Born effective charges~\cite{vanAken04p164} (the charges are very close to their nominal values).
\bibitem{coeff} The coefficients in the free energy model obtained for bulk YMnO$_3$ in Ref.~\onlinecite{Fennie05p100103} are (in eV unit) $\alpha_{02} = 0.016 $, $\beta_{04} = 0.002 $, $\alpha_{20} = -1.279$, $\beta_{40} = 0.813$, $\beta_{22} = 0.155$ and $\beta_{31} = -0.334$.
\end{thebibliography}
\end{document}